\documentclass[twocolumn, showpacs,amsmath,amssymb]{revtex4}
\usepackage{graphicx}
\usepackage{epstopdf}
\usepackage{bm}
\begin{document}
\preprint{09.2004}
\newcommand{\non}{\nonumber}
\newcommand{\be}{\begin{equation}}
\newcommand{\ee}{\end{equation}}
\newcommand{\bq}{\begin{eqnarray}}
\newcommand{\eq}{\end{eqnarray}}
\newcommand{\bsp}{\begin{split}}
\newcommand{\esp}{\end{split}}
\newcommand{\lps}{\langle}
\newcommand{\rps}{\rangle}
\title{Coherent transport of cold atoms in angle-tuned
optical lattices}
\date{\today}
\author{Roberto Franzosi}\email{franzosi@fi.infn.it}\affiliation{CNR-INFM UdR di Firenze,
Dipartimento di Fisica Universit\`a di Firenze,Via G. Sansone 1,
I-50019 Sesto Fiorentino, Italy.}
\author{Matteo Cristiani}
\author{Carlo Sias}
 \author{Ennio Arimondo}
 \affiliation{CNR-INFM UdR
di Pisa, Dipartimento di Fisica E. Fermi, Universit\`a di Pisa, Largo
Pontecorvo 3, I-56127 Pisa,
Italy.}
\begin{abstract}
  Optical lattices with a large
spacing between the minima of the optical potential can be created
using the angle-tuned geometry where the 1-D periodic potential is
generated by two propagating laser beams intersecting at an angle
 different from $\pi$. The present work analyzes the coherent transport
 for the case of this geometry.   We show that the potential depth can be kept constant
during the transport by choosing a \emph{magic} value for the laser
wavelength. This value agrees with that of the counterpropagating
laser case, and the magic wavelength does not  depend of the optical
lattice geometry. Moreover, we find that this scheme can be used to
implement controlled collision experiments under special geometric
conditions. Finally we study the transport of hyperfine-Zeeman
states of rubidium 87.

\end{abstract}
\maketitle
   \section{Introduction}
   Neutral atoms trapped in an artificial periodic potential formed by laser light, the so
called far detuned optical lattice, have been proposed as the
individual qubits for quantum information processing. In an optical
lattice,  neutral atoms can be trapped in the intensity maxima (or
minima) of a standing wave light field owing to the optical dipole
force.   A configuration with one single atom trapped in each site
of the optical lattice is realized  in the configuration of a
Mott-insulator transition associated to the loading of Bose-Einstein
condensates  (BECs) in optical lattices\cite{bloch05}. In order to
realize quantum gates with neutral atoms within the ideal
environment  of the Mott insulator several schemes have been
proposed. The common idea is to control the quantum atomic states
through the preparation and coherent manipulation of atomic
wave-packets by means of application of standard laser cooling and
spectroscopic techniques.  By using spin dependent, or more
precisely state dependent,  optical lattice potentials, the control
can be applied independently to multiple atomic  qubits based on
different internal states.  A state dependent potential may be
created for a one dimensional optical lattice in the so-called
lin-$\theta$-lin configuration, where the travelling laser beams
creating the optical lattice are linearly polarized with an angle
$\theta$ between their polarizations
\cite{finkelstein92,taieb93,marksteiner95}. In this configuration
the optical potential can be expressed as a superposition of two
independent optical lattices, acting on different atomic states. By
appropriately choosing the atomic internal states, the atoms will be
trapped by one of the two potentials depending on their internal
state. By changing the angle $\theta$ between the linear
polarizations of the two laser beams producing the optical lattice,
the wavepackets corresponding to orthogonal atomic states can be
coherently transported relative to each other
\cite{brennen99,jaksch99,jaksch05a}. Once the atomic qubits are
brought together they interact through controlled collisions. In the
coherent transport experiment of Mandel {\it et al} \cite{mandel03},
by a proper control of the angle $\theta$ the wavepacket of an atom
initially localized at a single lattice site was split into a
superposition of two separate wave packets, and delocalized in a
controlled and coherent way over a
defined number of lattice sites of the optical potential.\\
\indent In an optical lattice created by the counterpropagating
standing wave configuration, the spacing between neighboring minima
of the optical lattice potential is one half the wavelength of the
lasers creating the optical lattice.  Optical lattices with more
widely separated wells can be produced using long wavelength lasers,
as CO$_{2}$ lasers \cite{friebel98}. Alternatively, optical lattices
with a larger spacing between the minima/maxima of the optical
potential are formed using the angle-tuned geometry. There the
periodic potential is created by two laser beams propagating at an
angle $\phi$ and the lattice constant is
$d=\pi/\left[k\sin(\phi/2)\right]$,with $k=2 \pi/\lambda$ the laser
wavenumber \cite{morsch01,hadzibabic04,albiez05,fallani05}.\\
\indent The aim of the present work is to analyze the coherent
transport associated to the lin-$\theta$-lin polarization
configuration for the angle-tuned lattice geometry. We analyze
rubidium atoms in a given Zeeman level of a hyperfine state loaded
within a 1-D  optical-lattice. The 1-D geometry of the Bose gas may
be generated by a tight confinement along the orthogonal directions.
For instance a two-dimensional array of 1-D Bose gases (tubes) is
produced by confining the atoms through a two-dimensional optical
lattice generated by independent lasers,
as realized in Ref. \cite{moritz03}. \\
\indent Section II defines the geometry of the angle-tuned optical lattice. In Section III we analyze the
effective optical potential created by an  optical lattice in the angle-tuned configuration.
 The potential contains a component with a vectorial symmetry described through
 an effective magnetic field, as derived   in \cite{deutsch98}.
The dependence of the potential  depth on
the angles  defining the lattice geometry is analyzed in Section IV. Section V
 reexamines the coherent transport for the counterpropagating laser geometry.
 Section VI extends the coherent transport to the angle-tuned geometry and determines
 the condition for a constant optical lattice depth during transport. The application to the
 hyperfine-Zeeman states of rubidium is presented in Sec VII, and in the following Section the minimum time required
 to realize the coherent transport in the adiabatic limit is briefly discussed.\\

\section{Laser geometry}
\begin{figure}[ht]
\includegraphics[scale=0.7]{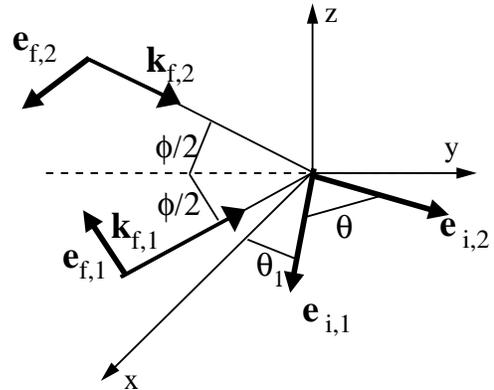}
\caption{Laser configuration determining the angle tuned geometry.
The laser fields propagate within the $(y,z)$ plane  with
wavevectors ${\bf k}^{\rm f}_{i}$ at an angle $\phi$ between them,
and polarizations ${\hat {\bf e}}^{\rm f}_{j}$,  with $(j=1,2)$.
These wavevectors and polarizations are generated applying the
rotations described in the text to two laser fields
counterpropagating along the $z$ axis  and with polarizations ${\hat
{\bf e}}^{\rm i} _{j}$,  with $(j=1,2)$ within the $(x,y)$ plane.}
\label{setup}
\end{figure}

The lasers generating the 1-D lin-$\theta$-lin
configuration are composed by two phase-correlated propagating
electric fields with frequency $\omega$ and amplitude $E_0$. Their
wavevectors
\begin{equation}
    \begin{split}
    {\bf k}^{\rm f}_{1} &= k(0,\cos\left(\phi/2\right),\sin\left(\phi/2\right)),\\
    {\bf k}^{\rm f}_{2} &= k(0,\cos\left(\phi/2\right),-\sin\left(\phi/2\right))
\end{split}
\end{equation}
lying on the $(y,z)$ plane create the angle-tuned geometry with
angle $\phi$, as shown in Fig. \ref{setup}.  The spatio-temporal
dependencies of the electric fields are
\begin{equation}
    {\bf {\cal E}}^{\rm f}_{j}({\bf x},t) ={\bf E}^{\rm f}_{j}({\bf x})e^{-i\omega t}+c.c.= \frac{E_{0}}{2} e^{ i\left( { {\bf
    k}^{\rm f}_{j} {\bf x}-\omega t}\right)} {
 \hat {\bf e}}^{\rm f}_{j} + c.c.\,  ,
 \label{electricfield}
\end{equation}
for $j=1,2$, with polarizations ${\bf \hat e}^{\rm f}_{j}$
defined in the following. This geometry creates
 a 1D optical lattice along the $z$-axis.
The laser fields   confining along the transverse directions are
not required for the
following analysis and not listed here.\\
\indent The above laser geometry is obtained by applying
 proper spatial rotations to a 1D optical lattice
initially created by two counterpropagating laser fields along the
$z$ axis. Let's introduce the rotation $R_{\rm x}(\alpha)$ of an
angle $\alpha$ around the $x$ axis
\begin{equation} R_{\rm x}(\alpha) =
    \begin{bmatrix} 1 & 0 & 0 \\
       0 & \cos\alpha & - \sin\alpha \\
       0 & \sin\alpha & \cos\alpha
    \end{bmatrix}
\label{rotx}
\end{equation}
and the rotation $ R_{\rm
z}(\alpha)$ of an angle $\alpha$ around the $z$ axis
\begin{equation}
R_{\rm z}(\alpha) =
 \begin{bmatrix} \cos\alpha & -\sin\alpha & 0 \\
\sin\alpha & \cos\alpha & 0 \\
0 & 0 & 1
\end{bmatrix} \, .
\label{rotz}
\end{equation}
In the lin-$\theta$-lin  counterpropagating
configuration the laser wavevectors are
\begin{equation}
    {\bf k}^{\rm i}_{1} = k(0,0,1), \quad
    {\bf k}^{\rm i}_{2} = k(0,0,-1),
\end{equation}
and  their  polarizations, ${\bf \hat e}^{\rm i}_{\rm j}$ for
$j=(1,2)$, are
\begin{equation}
    \begin{split}
    {\bf \hat e}^{\rm i}_{1} &= R_{\rm z} \left(\theta_1\right)
        \begin{bmatrix} 1  \\ 0 \\ 0 \end{bmatrix} \, , \\
    {\bf \hat e}^{\rm i}_{2} &= R_{\rm z}\left(\theta_1+\theta\right)
        \begin{bmatrix}1  \\ 0 \\ 0 \end{bmatrix}.
\end{split}
\label{polarization}
\end{equation}
Notice that{\bf ,} in addition to the angle $\theta$ between the two
polarization directions, we have introduced the angle $\theta_1$
between the polarization vector ${\bf \hat e}^{\rm i}_{1}$ and the
$x$ axis. The wavevectors  ${\bf k}^{\rm f}_{1,2}$, given by Eq. (1), are obtained
applying the following rotations:
\begin{equation}
{\bf k}^{\rm f}_{1}=R_{\rm x}(\frac {\phi - \pi}2){\bf k}^{\rm
i}_{1}\, ,\quad {\bf k}^{\rm f}_{2}=R_{\rm x}(\frac {\pi -
\phi}2){\bf k}^{\rm i}_{2}
\label{vectors}
\end{equation}
Such rotations are applied  as well to the polarization
vectors \cite{note1}
\begin{equation}
\begin{split}
{\bf\hat e}^{\rm f}_{1}&=R_{\rm x}(\frac{\phi - \pi}2){\bf\hat e}^{\rm i}_{1} \\
{\bf\hat e}^{\rm f}_{2}&=R_{\rm x}(\frac{\pi - \phi}2){\bf\hat e}^{\rm i}_{2}
 \label{polarizzazioni}
\end{split}
\end{equation}
The electric fields  ${\bf E}^{\rm f}_{1}({\bf x})$, ${\bf
E}^{\rm f}_{2}({\bf x})$  are obtained
by substituting Eqs. (\ref{vectors}) and (\ref{polarizzazioni}) into
 Eq. (\ref{electricfield}), and the total electric field
is given by
\begin{equation}
    {\bf E}_{\rm L}({\bf x}) = {\bf E}^{\rm f}_{1}({\bf x})+ {\bf
    E}^{\rm f}_{2}({\bf x})=\frac{E_{0}}{\sqrt{2}}{\bf e}_{\rm L}({\bf
    x})+c.c. ,
\label{totelectric}
\end{equation}
where ${\bf e}_{\rm L}({\bf x})$ defines the local polarization, not
necessarily unit norm.
\section{optical potential}
The optical potential experienced by the atoms is obtained
from the analysis of ref. \cite{deutsch98}. For alkali atoms, in the
limit that the laser detuning is much larger than the hyperfine
splittings in both the P$_{1/2}$ and P$_{3/2}$ excited states, the
optical potential has the following form \cite{note2}:
\begin{subequations}\label{EffPotD}
\begin{align}
    \hat{U}& = U_J \hat{I} + {\bf B}_{eff} \hat{\sigma}, \label{U} \\
U_J& =V^{0} \left|{\bf e}^2_L({\bf x})\right|, \label{UJ}\\
 {\bf B}_{eff}& = - iV^{1}\left[ {\bf e}^*_L({\bf x})
        \times {\bf e}_L({\bf x}) \right],
 \label{field}
\end{align}
\end{subequations}
where, in order to simplify the notation, we introduced the
following quantities:
\begin{equation}
\begin{split}
V^{0}(\omega)&=\frac{E_{0}^{2}}{2} \left(\frac{2\hat{\alpha}_{D2}}{3}+\frac{\hat{\alpha}_{\rm D1}}{3}
\right),\\
V^{1}(\omega)&=\frac{E_{0}^{2}}{2}\left(\frac{\hat{\alpha}_{D1}}{3}-\frac{\hat{\alpha}_{\rm D2}}{3}\right).
\end{split}
\end{equation}
representing the scalar and vector polarizabilities, derived for instance in \cite{park01}.
The operators $\hat{I}, \hat{ {\bf \sigma}}$ are the
identity and Pauli operators in the electron ground-state manifold.
The polarizabilities ${\tilde \alpha}_{D_{1}}$ and ${\tilde
\alpha}_{D_{2}}$ corresponding to the excitations to the
P$_{\frac{1}{2}}$ and P$_{\frac{3}{2}}$ excited states respectively,
depend on the dipole operator reduced matrix element
$ \lps J^\prime \Vert {\bf d} \Vert J=\frac{1}{2} \rps $ with
$J^\prime=\frac{1}{2},\frac{3}{2}${\bf :}
\begin{equation}
\label{alphas}
\begin{split}
{\tilde \alpha}_{D_{1}} = \frac{\vert \lps J^\prime =\frac{1}{2} \Vert {\bf d} \Vert J=\frac{1}{2}
\rps \vert^2}{\hbar \Delta_{D_{1}}} \, ,
\\
{\tilde \alpha}_{D_{2}} = \frac{\vert \lps J^\prime =\frac{3}{2} \Vert {\bf d} \Vert J=\frac{1}{2}
\rps \vert^2}{\hbar \Delta_{D_{2}}} \,.
\end{split}
\end{equation}
Here $\Delta_{D_{1,2}}$ is the detuning of the laser frequency
$\omega$ from the resonance between the states $\vert F_{max}=2\rps$
and $\vert F^{\prime}_{max}=2\rps$ or $\vert F^\prime_{max}=3\rps$,
for the D$_{1}$ or
D$_{2}$ lines of $^{87}$Rb  respectively.\\
\indent
The substitution of Eqs. (\ref{polarizzazioni}) for the local polarization in the right sides of
Eqs. (\ref{UJ}) and (\ref{field}) leads to
\begin{equation}
\begin{split}
U_J&=V^{0}\left[1+u(\theta_1,\theta,\phi)\cos(2\pi\frac{ z}{d})\right] \, , \\
{\bf e}_{\rm L}^{*}\times{\bf e}_{\rm L}&=i{\bf
b}(\theta_1,\theta,\phi) \sin(2\pi\frac{ z}{d})
\label{angledependence}
\end{split}
\end{equation}
where $d=\pi(k \sin{\frac{\phi}2})^{-1}$ is the period of the
optical lattice. The $u(\theta_1,\theta,\phi)$ parameter describes the spatial dependence of the scalar part of the optical
lattice
\begin{equation}
u(\theta_1,\theta,\phi)=\cos(\theta +
2\theta_{1})\cos^2(\frac{\phi}2)+\cos(\theta) \sin^2(\frac{\phi}2) \, .
\end{equation}
The spatial components of the effective magnetic field are given by
\begin{equation}
\begin{split}
b_{\rm x}(\theta_1,\theta,\phi)&=-\sin(\theta_{1})\sin(\theta_1+\theta)\sin(\phi), \\
b_{\rm y}(\theta_1,\theta,\phi)&= \sin(2\theta_1 + \theta) \cos(\frac{\phi}2), \\
b_{\rm z}(\theta_1,\theta,\phi)&=-\sin(\theta)\sin(\frac{\phi}2),
\end{split}
\end{equation}
determining also the module $b=|{\bf b}|$. $u$ and $b$
satisfy the following useful relations:
\begin{equation}
    \label{useful1}
        u^2(\theta_1, \theta, \phi) + b^2(\theta_1, \theta, \phi) = 1,
\end{equation}
and
\begin{equation}
    \label{useful2}
        u(\theta_1 = 0, \theta, \phi) = \cos( \theta ).
\end{equation}
 \indent  The effective
magnetic field ${\bf B}_{eff}$ \cite{note3}  varies spatially
with a $d$ period. Its components along the three axes
have amplitudes depending on the angles defining the lattice
geometry. If the light field is everywhere linearly polarized,
$\theta_1=\theta=0$, the effective magnetic field vanishes and the
light shift is independent of the magnetic atomic sublevel{\bf :}
${\hat U}({\bf x}) = U_{J}({\bf x}){\hat I}$.   For the
counterpropagating geometry, i.e. $\phi =\pi$, investigated by
\cite{brennen99,jaksch99} and implemented in \cite{mandel03}, the
effective magnetic field is oriented along $z$-axis.

\section{Hamiltonian Eigenvalues}
Making the assumption of neglecting the kinetic energy of the atoms,
the effective potential corresponds to the full hamiltonian acting
on the atomic states, and the position $z$ can be treated
as an external parameter.  If we consider the
two-dimensional subspace characterized by the electron spin
component, {\it i.e.}, $|S=\frac{1}2,m_{S}=\frac{1}2 \rangle$ and
$|S=\frac{1}2,m_{S}=-\frac{1}2 \rangle$,  the eigenvalues of the
hamiltonian are
\begin{equation}
\begin{split}
\epsilon_{+}&=V^{0}u \cos(2\pi  \frac{z}{d})+V^{1}b \sin(2\pi  \frac{z}{d}), \\
\epsilon_{-}&=V^{0}u \cos(2\pi  \frac{z}{d})-V^{1}b \sin(2\pi \frac{z}{d}),\\
\end{split}
\label{eigenvalue}
\end{equation}
with a constant term  $V^{0}$ left out. These quantities represent
the optical potential experienced by the atoms. In an equivalent
description, $\epsilon_{\pm}$ define the energies of the
$|S,m_S=\pm\frac{1}{2}\rangle$ atomic states when the electron
spin is aligned along the local direction of the magnetic
field, {\it i.e.}, $\hat {\sigma}=\pm{\bf b}/b$. In fact we write
\begin{equation}
\begin{split}
\epsilon_{+}&= U_J + |{\bf B}_{eff}\ | ,\\
\epsilon_{-}&= U_J - |{\bf B}_{eff}|.\\
\end{split}
\label{spinenergies}
\end{equation}
\indent The eigenvalues $\epsilon_{\pm}$   of Eqs (\ref{eigenvalue})  can be expressed as
\begin{equation}
  \begin{split}
      \epsilon_{+}(\theta_1,\theta,\phi,z)&=U_{0} \cos(2\pi \frac{z}{d}+\gamma_0),\\
      \epsilon_{-}(\theta_1,\theta,\phi,z)&=U_{0} \cos(2\pi \frac{z}{d}-\gamma_0). \
  \end{split}
\label{eigen2}
\end{equation}
The potential depth  $U_{0}(\theta_1,\theta,\phi)$ and the  relative phase $2\gamma_0(\theta_1,\theta,\phi)$ are given by
\begin{subequations}
\label{generic}
\begin{align}
U_{0}&=V^{0}\sqrt{\eta^{2}+(1-\eta^{2})u^{2}},\label{generica}\\
\gamma_0&=-\arctan\left(\eta\sqrt{ \frac{1}{u^{2}}-1}\right),\label{genericb}
\end{align}
\end{subequations}
where we have introduced
\begin{equation}
\eta(\omega)=\frac{V^{1}(\omega)}{V^{0}(\omega)}.
\end{equation}
Eq. (\ref{eigen2}) demonstrates the spatial periodicity of the
potentials experienced by the atomic eigenstates. Ultracold atoms
are trapped at the spatial positions corresponding to the minima of
the optical lattice potentials. For positive $U_0${\bf ,} the $z_+$
and $z_-$ minima positions  for the  $|S=\frac{1}2,m_{S}=\frac{1}2
\rangle$ and $|S=\frac{1}2,m_{S}=-\frac{1}2 \rangle$ states
respectively are given by
\begin{equation}
\begin{split}
\frac{z_+}{d}&=\frac{1}{2}\left(1-\frac{\gamma_0}{\pi}\right) +l \, , \\
\frac{z_-}{d}&=\frac{1}{2}\left(1+\frac{\gamma_0}{\pi}\right)+l\, ,
\end{split}
\end{equation}
with $l$ an integer. For instance at $\gamma_0=\pi/2$, the $|+\rangle$
atoms are localized at $z_+=(l+\frac{1}{4})d$ and the $|-\rangle$ atoms at
$z_-=(l+\frac{3}{4})d$. At  $\gamma_0=0$ both species are localized at
the $(l+\frac{1}{2})d$ positions.

\section{Counterpropagating geometry}
We consider the case $\phi=\pi$ of an optical lattice created by two
counterpropagating laser fields. Then  the functions $u$ and $b_j$ determining the optical lattice
potential reduce to
\begin{equation}
    \begin{split}
        &u= \cos( \theta )
        \\
        &b_x= b_y= 0
        \\
        &b_z = - \sin( \theta )
    \end{split}
\end{equation}
For this geometry the functions $u$ and $b_j$  do not
depend on the angle $\theta_1$ but only on the relative angle
$\theta$. This is a consequence of the symmetry of the system. Since
the two beams forming the optical lattice propagate along the
direction $z$, the
system is invariant under rotations around that axis.\\
\indent At $\phi = \pi$ the potential depth $U_0(\theta)$ and the phase shift $\gamma_0(\theta)$ become
\begin{equation}
    \label{counter}
    \begin{split}
        U_0(\theta) &= V^{0} \sqrt{ \eta^2 + ( 1 - \eta^2 ) \cos^2( \theta )}
        \\
        \gamma_0(\theta) &=- \arctan\left( \eta \tan( \theta )\right)
        \\
    \end{split}
\end{equation}
\begin{figure}[ht]
\includegraphics[width=1.0\linewidth]{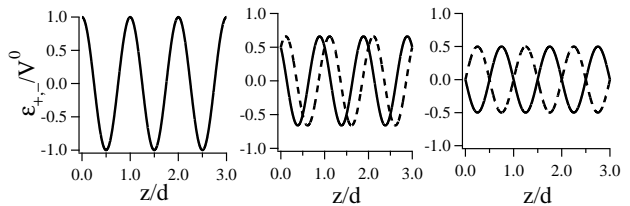}
\caption{Eigenvalues $\epsilon_+$ (continuous line) and $\epsilon_-$
(dashed line), in units of $V^{0}$,  plotted for different values of
$\theta$ ($0$, $\pi/3$ and $\pi/2$ from left to right) in the case
of the counterpropagating geometry corresponding to $\phi=\pi$ and
for the value $\eta=-0.5$. } \label{fig_2}
\end{figure}
Fig. (\ref{fig_2}) reports the two eigenvalues of Eqs.
(\ref{eigen2}) as a function of $z$ for different values of the
relative angle $\theta$ between the two linear
polarizations. The laser frequency is chosen such  that $V^{1}=-0.5 V^{0}$, that is
$\eta=-0.5$.  At $\theta=0$ with  the laser polarizations parallel,
$\epsilon_+$ and $\epsilon_-$  coincide. By increasing $|\theta|$,
 the minima of the
potential curves $\epsilon_+$ and $\epsilon_-$ move in opposite
directions along the $z$ axis, and the potential depth
decreases. At $\theta=\pi/2$ the minima of $\epsilon_+$ coincide with
the maxima of $\epsilon_-$, and their amplitudes are at the minimum.
 Let us suppose to start at $\theta=\pi/2$ preparing a $|+\rangle$ atom
 at the site $id$ and a $|-\rangle$ atom  at the site $(i-\frac{1}{2})d$.
Varying adiabatically $\theta$ from $\pi/2$ to $0$  (or to $\pi$) the two
particle will occupy the same site. This protocol was
used to transport the atoms from one site to the other in order
to produce controlled collisions \cite{brennen99,jaksch99,mandel03}.
We recall that in this counterpropagating geometry the effective magnetic
field is oriented along the $z$ axis and is equal to zero for $\theta=0$, when the atoms collide.\\
\indent  Fig. \ref{fig_3} shows that the potential depth does
not remain constant by varying $\theta$. As pointed in refs
\cite{jaksch99,jaksch05a}, this difficulty is avoided for a
particular choice of the parameter $\eta$. In fact for $\eta^2=1$,
the term $\cos^2\theta$ disappear in Eq. (\ref{counter}) and $U_0$
becomes a constant independent of $\theta$ and equal to $V^{0}$.
\begin{figure}[ht]
\includegraphics[width=0.9\linewidth]{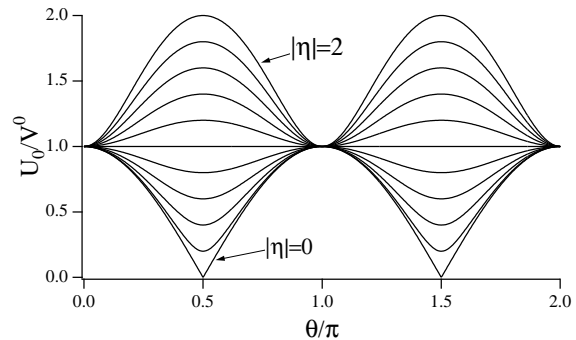}
\caption{Potential depth $U_0$ as a function of $\theta$ for
$|\eta|$ ranging between $0$ and $2$ in steps of $0.2$, from bottom
to top. For $|\eta|=1$ the potential is a constant  independent of
$\theta$.} \label{fig_3}
\end{figure}

\begin{figure}[ht]
\includegraphics[width=0.9\linewidth]{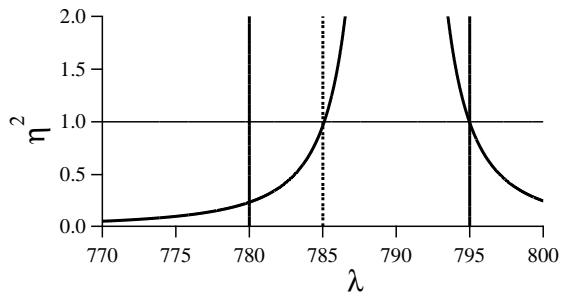}
\caption{Plot of the dimensionless parameter $\eta^2$ as a function
of the laser wavelength $\lambda$. The two continuous vertical lines
denote the position of the resonant wavelengths $\lambda_{D2}$ and
$\lambda_{D1}$, respectively. The dashed vertical line indicates the
magic wavelength $\lambda^*$.} \label{fig4}
\end{figure}

 By assuming equal   dipole moments for the D$_1$ and D$_2$ lines
 $\left| \langle J'=\frac{1}{2} \left\| {\bf d}
\right\| J=\frac{1}{2} \rangle\right|^2  = \left| \langle
J'=\frac{3}{2}
 \left\| {\bf d} \right\| J=\frac{1}{2} \rangle\right|^2$\cite{note4}, the parameter $\eta(\omega)$
becomes{\bf :}
\begin{equation}
    \eta(\omega) = \frac{ \Delta_{D2}- \Delta_{D1} }{
    \Delta_{D2} +2\Delta_{D1}}.
\end{equation}
Fig. \ref{fig4} reports the parameter $\eta$ versus the
laser wavelength $\lambda$. The
constraint $\eta^2 = 1$ implies $\eta = 1$ or $\eta = -1$. The
first condition is satisfied for $\Delta_{D1} = 0$ which is not an
acceptable value, since the whole treatment for the optical lattice
potential is valid only for detunings  $\Delta_{D1,2}$ large with
respect to the typical hyperfine splitting. The $\eta = -1$ relation
is satisfied if the laser wavelength is equal to the \emph{magic}
value $\lambda^*$
\begin{equation}
    \frac{1}{\lambda^*} = \frac{1}{3} \left( \frac{1}{\lambda_{D1}} +
    \frac{2}{\lambda_{D2}} \right)
\label{magic}
\end{equation}
where $\lambda_{D1}$ and $\lambda_{D2}$ denote the resonant
wavelengths for the D$_1$ and D$_2$ lines. For $\eta = -1$ the
relative phase $\gamma_0$ becomes
\begin{equation}
\begin{split}
\gamma_0&=\theta\quad\quad\quad\quad\, for\,\, 0<\theta<\pi
\\ \gamma_0&=2\pi - \theta\quad\quad for\,\, \pi\leq\theta\leq2\pi
\end{split}
\end{equation}
Therefore for the magic wavelength {\bf $\lambda^*$} the potential depth is
independent on $\theta$ and the phase $\gamma_0$ varies linearly with
the relative angle $\theta$.
\begin{figure}[ht]
\includegraphics[width=1.0\linewidth]{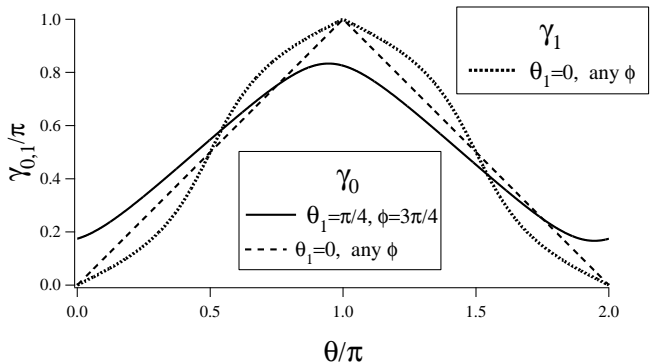}
\caption{Optical lattice phase shifts $\gamma_0$ and $\gamma_1$
versus $\theta$ for the laser at the magic wavelength, corresponding
to $\eta=-1$.  For $\gamma_0$ the continuous line corresponds to
$\theta_1=\pi/4$ and $\phi=3\pi/4$.  The dashed line plots
$\gamma_0$ at  $\theta_1=0$ whichever value of $\phi$.  The same
$\gamma_0$ values are obtained for $\phi=\pi$ whichever value of
$\theta_1$. For $\gamma_1$  the dotted line corresponds to
$\theta_1=0$ whichever value of $\phi$. The coherent transport
requires for the phase shift $\pi/2$ as initial value and $0$ as
final value. A complete transport cannot be realized for the
conditions corresponding to the continuous line, but instead  can be
realized for the dashed and dotted lines.} \label{fig_5}
\end{figure}

\section{angle-tuned configuration}
For  $\phi$ different from $\pi$ the potential depth $U_0$ and the
phase $\gamma_0$ depend on the angle $\theta_1$ as well. This reflects
the fact that for $\phi \ne \pi$ the system is not invariant under
rotations around the $z$ axis.  \\
\indent For the $U_0$ dependence on
$\theta_1$   the magic
wavelength $\lambda^*$ plays again a key role. In fact
 by choosing
$\eta =- 1$ Eqs. (\ref{generic}) reduce to
\begin{equation}
 \label{eqangletuned}
   \begin{split}
   U_0 &= V^{0}\;, \\
 \gamma_0(\theta_1,\theta,\phi) &= \arctan{\sqrt{\frac{1}{u^2}-1}}.
    \end{split}
   \end{equation}
Therefore the potential depth $U_0$ depends only on the laser wavelength. \\
\indent The remaining dependence of $\gamma_0$ on the angles
$(\theta_1,\theta,\phi)$ defining the optical lattice represents a
difficulty for the coherent transport operation. For instance for
the values of $\theta_1=\pi/4$ and $\phi=3\pi/4$ the phase shift
$\gamma_0$ versus $\theta$ is plotted as a dashed line in Fig. \ref{fig_5} and is
compared to the value corresponding to the  counterpropagating geometry, plotted
as a continuous line. For those
values of $\phi$ and $\theta_1$ the range of variation of $\gamma$
is smaller than $\pi$. This means that by varying $\theta$ the two
potential curves corresponding to the two atomic eigenstates
are shifted by a quantity smaller than the spatial period $d$. Thus
the minima of the potentials for the two atomic eigenstates do
not coincide and a complete transport is not realized. However
under the special condition of $\theta_1 = 0$  Eq.
(\ref{useful2}) holds, and the phase $\gamma_0$ becomes
\begin{equation}
\begin{split}
\gamma_0(\theta)&=\theta\quad\quad\quad\quad\, for\,\, 0<\theta<\pi,\\
\gamma_0(\theta)&=2\pi - \theta\quad\quad for\,\, \pi\leq\theta\leq2\pi.
\end{split}
\end{equation}
Therefore by choosing $\lambda= \lambda^*$ and $\theta_1=0$, the phase
shift becomes fully equivalent to that of the counterpropagating case. Thus, it is
possible to move the two potential curves of Eqs.  (\ref{eigen2})
without changing the potential depth, and  a
spatial coherent transport with amplitude $d$ can be produced by
varying the angle $\theta$ of the laser polarizations.  For this angle-tuned
configuration, even if the effective magnetic field is null  at $\theta=0$ where
 the atoms collide,  it includes a component along the $y$ axis at other values of
 $\theta$.

\section{Transport of rubidium states}
\indent The vector component and the total potential of Eqs.
(\ref{EffPotD}) experienced by  the atoms depend on the projection
of the electron magnetic moment or, equivalently, of the total
angular momentum ${\bf \hat{F}}$ along the local magnetic field.
Different effective potentials are experienced by different
Zeeman levels of the ground hyperfine state, and the laser
parameters required for the coherent control depend on the atomic
computational basis. Here we consider two different hyperfine-Zeeman
states $|F,m_F\rangle$ of the $^{87}$Rb atom as in the analysis of
\cite{brennen99,jaksch99,mandel03}. As described in \cite{jaksch99}
the potential experienced by the atoms in these internal states is
derived from the potential of Eq. (\ref{U}) by considering the
components of the electron spin for these states. However in the
angle tuned geometry with $\phi \ne \pi$, the effective magnetic
field also includes components oriented along the $x$ and $y$ axes.
As a consequence the potential experienced by the atomic states
depend on the optical lattice loading process and the occupation
amplitudes for the Zeeman states. In fact Raman coherences of the
type $|F,m\rps \to |F^{\prime},m\pm 1\rps$ are created by the
effective magnetic field.  As pointed out in
ref. \cite{deutsch97}, in the tight binding regime where each lattice
site can be considered as an independent potential well, the Wannier
states constituting an orthonormal basis within each well in general
become spinors. \\
\indent For simplicity we analyze the case of an
adiabatic loading of the $|F,m_F\rangle$ states in the lattice so
that their $m_F$ component is oriented along the local magnetic
field $ {\bf B}_{\mathrm{eff}}$.
Thus we impose the atomic states under consideration to be the $|F,m_F\rangle$ eigenstates along the local magnetic field. Let's consider the following states:
\begin{widetext}
\begin{equation}
\label{states}
    \begin{split}
    |0>&=|F=2,m_F=2\rangle
          =|I=\frac{3}{2},m_I=\frac{3}{2}\rangle|S,m_S=\frac{1}{2}\rangle\, ,\\
    |1>&=|F,m_F\rangle \,
    =c_+|I=\frac{3}{2},m_I=m_F-\frac{1}{2} \rangle |S,m_S=\frac{1}{2}\rangle+c_-|I=\frac{3}{2},m^\prime_I =m_F+\frac{1}{2}\rangle |S,m_S=-\frac{1}{2}\rangle \, ,\\
     \end{split}
\end{equation}
the coefficients $c_{\pm}$ defining the normalized superposition.  For the $|F,m_F=1\rangle$ state of the explorations in refs. \cite{brennen99,jaksch99,mandel03} the coefficients are $c_+=1/2$ and $c_-=\sqrt{3}/2$.
 The energies $E_{0,1}$ of
the states $|0,1\rangle$ at a fixed position $z$ within
the optical lattice are given by
\begin{equation}
\begin{split}
E_0\left(\theta_1,\theta,\phi,z\right)&=\epsilon_+\left(\theta_1,\theta,\phi,z\right)=U_{0} \cos(2\pi \frac{z}{d}+\gamma_0) \, , \\
E_1\left(z,\theta_1,\theta,\phi\right)&=|c_+|^2 \epsilon_+ \left(\theta_1,\theta,\phi,z\right)
+|c_-|^2\epsilon_-\left(\theta_1,\theta,\phi,z\right)
= V^0u \cos(2\pi  \frac{z}{d})+V^1b\left(|c_+|^2-|c_-|^2\right) \sin(2\pi  \frac{z}{d})\\
&=U_{1} \cos(2\pi \frac{z}{d}-\gamma_1).\\
\end{split}
\end{equation}
\end{widetext}
While $U_0$ and $\gamma_0$ are given by Eq. (\ref{generic}), the potential depth  $U_{1}(\theta_1,\theta,\phi)$ and the phase
$\gamma_1(\theta_1,\theta,\phi)$ are given by
\begin{subequations}
\label{Fpotentialphase}
\begin{align}
U_{1}&=V^{0}\sqrt{\eta^{2}\Delta^2_c+(1-\eta^{2}\Delta^2_c)u^{2}}, \label{U1}\\
\gamma_1&=-\arctan\left(\eta \Delta_c\sqrt{ \frac{1}{u^{2}}-1}\right)\, ,\label{gamma1}
\end{align}
\end{subequations}
where
\begin{equation}
\Delta_c=|c_-|^2-|c_+|^2.
\end{equation}
We obtain two different effective  lattice
potentials trapping the atoms in the$|0 \rangle$ and $|1 \rangle$ hyperfine-Zeeman
states.\\
\indent Because Eqs. (\ref{Fpotentialphase}) have the same structure as Eqs.  (\ref{eqangletuned}),
the coherent transport is determined by the $u(\theta)$ dependence at fixed
values $\theta_1$ and $\phi$.   By using this analogy we conclude that, in order to perform a
controlled-collisions experiment, the potentials seen by the two
hyperfine states $|0,1 \rangle$ must move in opposite
direction when $\theta$ is varied. The comparison Eq. (\ref{gamma1}) to Eq. (\ref{genericb}) indicates
 that this condition is satisfied when
the $|1 \rangle$ state is chosen such that $\Delta_{c} > 0$, that is, when
$|c_{-}|
> |c_{+}|$. For the states $|F=2,m_{F}=-2 \rangle$, $|F=2,m_{F}=-1
\rangle$, $|F=1,m_{F}=1 \rangle$
this inequality is satisfied.

\indent The optimal coherent transport is obtained when the potential depth
 is constant by varying the $\theta$ control parameter. In Section V
 the constance of the optical depth was realized by fixing the laser wavelength
 at the magic value. For the present case of two hyperfine states a unique magic
 wavelength where the $U_0$ and $U_1$ potential depths are both independent
 of $\theta$ does not exist.   While the $U_0$ constance imposes $\eta^2=1$ and
 produces the magic wavelength Eq. (\ref{magic}), the $U_1$ constance imposes $\left(\eta\Delta_c\right)^2=1$ leading to a different laser wavelength.  For instance,
 at $\theta_1=0$ and fixing  the laser wavelength to the
$\lambda^{*}$ value of Eq. (\ref{magic}) such that $\eta=-1$,  the potential depths for
the $|0,1 \rangle$ states become
\begin{equation}
\begin{split}
U_0&=V^{0} \, , \\
U_1(\theta)&=V^{0}\sqrt{\Delta_c^2+(1-\Delta_c^2)\cos^{2}\theta},
\end{split}
\end{equation}
while their phases $\gamma_{0,1}$ are
\begin{equation}
\begin{split}
\gamma_0(\theta)&=\theta\, , \\
\gamma_1(\theta)&=\arctan\left(\Delta_c\tan\theta\right).
\end{split}
\end{equation}
When $\theta$ is varied from $\frac{\pi}{2}$
to $0$ or $\pi$, while the $U_0$ potential depth  is independent of
$\theta$,  the $U_1$ potential depth depends on $\theta$ and its range
is determined by $\Delta_c$, whence by the $c_\pm$ values. Therefore, when $\theta$ is varied the potentials experienced
by the $|0,1\rangle$ states move with different velocities. The phase
$\gamma_0$ is linearly dependent of $\theta$, while the
dependence of $\gamma_1$ on $\theta$ has a more complicated
behavior.  For the case of the $|F=1,m_{F}=1 \rangle$ state the $\gamma_1$
dependence   is shown  by the dotted line
in Fig. \ref{fig_5}.  The coherent transport condition of phase shifts $\gamma_{0,1}$ varying from $\pi/2$ to $0$ is realized for both the hyperfine-Zeeman states. Different results for the change in the potential depth and for
the phase dependence on $\theta$, and therefore for the displacements of the two
potentials, are obtained for a laser wavelength different from the magic one.\\

\section{Adiabatic transport}
In order to realize an efficient quantum gate, the time dependence
of polarization angle $\theta$ should be chosen so that the
transport of the atomic states is realized in the adiabatic limit,
{\it i.e.} the atoms remain in the ground state of local optical
potential. For an analysis of the adiabaticity constraint necessary
for the coherent transport we approximate the atomic potential of
Eq. (\ref{eigen2}) with a harmonic one.   For non interacting atoms
experiencing a harmonic potential moving with respect to the
laboratory frame, the Hamiltonian may be written as
\cite{salomon97}
\begin{equation}
H=\frac{p^2}{2m}+\frac{1}{2}m\Omega^2z^2+M\ddot{z}_{hp}(t)z,
\end{equation} where
$z$ denotes the coordinate of the atom in the harmonic potential frame,
$\ddot{z}_{hp}$ represents the acceleration of the harmonic potential
in the laboratory frame, and the oscillation frequency $\Omega$ of the harmonic oscillator is
\begin{equation}
\Omega=\sqrt{\frac{U_0}{M}}\frac{4 \pi}{\lambda}sin(\phi /2).
\end{equation}
If we consider the last term of the above Hamiltonian as
a time-dependent perturbation, the probability of transferring an atom
to the first excited level of the harmonic potential is
\begin{equation}
P_{f,i}=\frac{1}{2a_{ho}^2\Omega^2}\left|\int_{0}^{T}\ddot{z}_{hp}(t)e^{i\Omega
t}dt\right|^{2},
\label{probability}
\end{equation}
where $a_{ho}=\sqrt{\hbar/M\omega}$ is the ground state radius for the harmonic
potential, and $T$ is the time required for the coherent transport process. \\
\indent The amplitude of the transfer probability of Eq. (\ref{probability}) depends greatly
on the time dependence of the $\ddot{z}_{hp}$ acceleration.
At first we will suppose  that, as in the theoretical analysis of
ref. \cite{jaksch99} and in the experimental investigation of ref.\cite{mandel03},
  the potential moves at a constant speed, by
imposing an infinite acceleration at $t=0$ and $t=T$.
For this transformation the adiabatic condition for the atomic transformation requires
\begin{equation}
T\gg\frac{d}{a_{ho}}\frac{1}{\Omega}=\frac1{2}\left(\frac{\lambda^{6}M^3}{(4\pi)^2\hbar^2U_0}\right)^{1/4}
sin\left(\phi/2\right)^{-\frac{3}{2}}.
 \label{adiabaticlimit}
\end{equation}
Owing to this dependence on $\phi$, for a constant speed of the potential the time
$T$ required to realize an adiabatic coherent transport in the angle-tuned configuration
is much longer than in the counterpropagating case, for a given depth of the optical
potential . Such result could impose a strong constraint for  performing quantum computation
 with angle-tuned lattices. \\
 \indent However, the condition on $T$ for realizing the
 adiabatic limit becomes less restrictive if we assume a different motion for
the lattice harmonic potential confining the atoms. For instance,
let us assume that the lattice is constantly accelerated from $t=0$
to $t=\frac T{2}$ and constantly decelerated from $t=\frac T{2}$ to
$t=T$. Thus  the adiabatic condition becomes
\begin{equation}
T\gg\sqrt{\frac{d}{a_{ho}}}\frac{1}{\Omega}=\frac1{\sqrt{2}}\left(\frac{\lambda^{10}M^5}{(4\pi)^6\hbar^2U_0^3}\right)^{1/8}
sin\left(\phi/2\right)^{-\frac5{4}}.
 \label{adiabaticlimit2}
\end{equation}
Therefore using this motion of the potential,  the dependence of the
minimum time for the coherent transport  on the  angle $\phi$ is
modified causing a decrease in the time by a factor which for
$\phi\sim5^\circ$ is larger than 2. Moreover the additional
dependence on $\lambda$, $U_0$, and the physical constants of Eq.
(\ref{adiabaticlimit2}) contributes to the decrease of the time
scale, so that for the experimental conditions of ref.
\cite{mandel03} transport times around 10 $\mu$s can be achieved.
\section{Conclusions}
In quantum computation experiments with neutral atoms loaded in
optical lattices, a crucial aspect is the single site
addressability. In the angle-tuned configuration where the   lattice
constant could be  large, the question of the single site
addressability is shifted to a frame of more accessible dimensions.
For this geometry  the coherent transport protocol requires specific
conditions of  the laser beam polarizations,  linked to the breaking
of the rotational symmetry associated to the counterpropagating
geometry.  An additional request   is the constance of the  optical
potential depth during the coherent transport. This constance is
realized    by choosing a \emph{magic} wavelength for the laser
fields producing the lattice.  The value of the magic wavelength is
independent of the lattice geometry. However an unique magic
wavelength for the transport of all hyperfine-Zeeman atomic states
does not exist.

Coherent transport within an optical lattice represents a component of the
process based on  ultracold collisions and leading to entanglement of  neutral atoms
and implementation of quantum logic.  By storing the ultracold atoms in the microscopic
potentials provided by optical lattices the collisional interactions can be controlled via
laser parameters. At the low temperature  associated to the Mott insulator, the
collisional process is described through s-wave scattering. In the $\theta_1=0$
laser configuration of the angle-tuned geometry,  at  $\theta=0$  the effective magnetic field is
null and the scattering potentials associated to the  different
atomic states have an identical spatial dependence. However, as new feature brought  by the
angle-tuned geometry, at $\theta \ne 0$
the effective magnetic field is different from zero and oriented along different directions
for different hyperfine-Zeeman states. Therefore during the whole collisional process the
colliding atoms may be oriented along different spatial directions.  In order to treat this
collisional configuration,  the atomic interaction  may be described
through the pseudopotential models introduced in refs. \cite{bolda03,stock05}
for asymmetric trap geometries.

The atomic control is based on  a the realization of a  Mott
 insulator phase, in which the number of atoms occupying each lattice site is fixed.
 The physics of such a system is described in terms of a Bose-Hubbard
model whose Hamiltonian  contains the on-site repulsion resulting from
the collisional interactions between the atoms, and the
hopping matrix elements that take into account the tunneling rate of the
atoms between neighboring sites. Both the repulsive interaction and
the hopping energy can be tuned by adjusting the
lasers setup, as reviewed in \cite{jaksch05b}. The Mott insulator phase
 is realized under precise conditions between the
 on-site repulsion and the hopping matrix elements.
 The dependence of these parameters defining the angle-tuned lattice should
 be investigate in order to realize a Mott insulator in an angle tuned geometry.

\section{ Acknowledgments }
This researach was financially supported by the EU through the STREP Project OLAQUI and by the Italian MIUR through a PRIN Project. The authors are gratefully to Dieter Jaksch and Carl J. Williams for useful discussions.



\begin{thebibliography}{99}
\bibitem{note1} Owing to this transformation the angle between the
polarizations of the lin-$\theta$-lin geometry is
$\cos^{2}\left(\theta/2\right)+ \sin^{2}\left(\theta/2\right) \cos \phi.$
\bibitem{note2} The atomic motion is determined also by the
gravitational and trapping potentials not included here.
\bibitem{note4} On the basis of the lifetime measurements of \cite{volz96} the ratio of the dipole moments for the D$_2$ and D$_1$   is 1.027(1).
\end{thebibliography}
\end{document}